\documentclass[sigplan,screen,nonacm]{acmart}
\AtBeginDocument{
  }

\setcopyright{none}
\renewcommand\footnotetextcopyrightpermission[1]{}
\usepackage[table]{xcolor}
\usepackage{amsmath}
\usepackage{pifont}
\usepackage{listings}
\usepackage{multirow}
\usepackage{xspace}
\usepackage{graphicx}
\usepackage{booktabs}
\usepackage{enumitem}
\usepackage{hyperref}
\usepackage{stmaryrd}

\newcommand{\skyegg}{\texttt{SkyEgg}\xspace}
\definecolor{brickred}{rgb}{0.8, 0.25, 0.33}
\definecolor{darkspringgreen}{rgb}{0.09, 0.45, 0.27}

\newcommand{\x}{\texttimes\xspace}

\newcommand{\dhalfcheck}{\textcolor{darkspringgreen}{\ding{51}}\textsuperscript{\textcolor{brickred}{\kern-0.5em\tiny\ding{55}}}}
\newcommand{\dcheck}{\textcolor{darkspringgreen}{\ding{51}}}

\newcommand{\boldparagraph}[1]{\emph{{\textbf{#1}}}\xspace}
\DeclareMathOperator*{\argmin}{arg\,min}
\newcommand{\aggressivebreak}{\allowdisplaybreaks[4]\interdisplaylinepenalty=0}

\aggressivebreak

\begin{document}

\title{SkyEgg: Heterogeneity-Aware Hardware Synthesis via Equality Saturation}

\author{Yuyang Zou}
\affiliation{\institution{Peking University}\country{China}}
\author{Youwei Xiao}
\affiliation{\institution{Peking University}\country{China}}
\author{Yitian Sun}
\affiliation{\institution{Peking University}\country{China}}
\author{Yun Liang}
\affiliation{\institution{Peking University}\country{China}}

\begin{abstract}
Hardware synthesis is a key interface between high-level programs and accelerator designs.
Modern FPGAs increasingly expose heterogeneity in resource functionality and timing configurability. Exploiting these resources requires synthesis to choose among hardware implementations and fine-grained configuration options, and coordinate these choices across multiple stages.
However, prior frameworks fail to fully explore these choices due to resource models confined to per-operation mappings and limited fusion patterns.
Moreover, they perform transformation, scheduling, and mapping sequentially, selecting an expression form and assigning cycle boundaries before full mapping information is available, both of which preclude chaining or fusion and limit the resource choices.
Consequently, viable implementations are excluded from subsequent optimization stages, limiting the end-to-end performance potential of the synthesis flow.

We present \skyegg, a heterogeneity-aware hardware synthesis framework for FPGAs. The key insight of \skyegg is to bring algebraic equivalence and diverse resource capabilities together throughout the synthesis flow.
Specifically, it encodes target-specific hardware mappings and their configurations as extensible mapping rules alongside algebraic rewrite rules in equality saturation, placing algebraic and resource alternatives in a common representation.
It further proposes a novel transformation- and mapping-aware scheduling formulation that uses these alternatives and their configuration-specific timing to jointly select an equivalent form, the resource mappings it enables, and cycle assignments to minimize end-to-end latency under target-frequency and hardware constraints, thereby unlocking the hidden performance potential of target resources.
\skyegg also provides a heuristic solver that rapidly finds high-quality solutions for large designs. Against two state-of-the-art commercial HLS tools, \skyegg achieves average speedups of 3.12\x over Vitis HLS and 3.18\x over Altera HLS across target frequencies.
\end{abstract}

\maketitle

\section{Introduction}

Hardware synthesis forms a critical interface between high-level programs and accelerator designs~\cite{cong_high-level_2011,xu_hector_2022,josipovic_dynamically_2018,ferrandi_invited_2021,canis_legup_2013,amd_inc_vitis_2025-1,wu_graphhls_2026}.
It translates computations into hardware implementations by determining how operations are mapped, organized, and executed, which can substantially affect accelerator performance.
This task becomes increasingly challenging as target platforms incorporate diverse, specialized, and configurable hardware resources.
Modern FPGAs exemplify this trend by integrating programmable logic with configurable Digital Signal Processor (DSP) blocks~\cite{amd_inc_ultrascale_2021,intel_intel_2025} and parameterized, architecture-specific Intellectual Property (IP) cores capable of implementing fused arithmetic patterns~\cite{xilinx_inc_floating-point_2020}.
These resources create substantial performance opportunities, making their effective use a central task for hardware synthesis.

The challenge arises from the heterogeneity of hardware resources in both functional capability and timing behavior.
Functionally, a configurable resource may support diverse computation patterns, yet an input expression may not directly match any of them, requiring \emph{transformation} to expose a form that admits a compatible \emph{mapping}.
Temporally, resource types and configurations implementing the same functionality may have different cycle latencies and timing characteristics.
\emph{Scheduling} must account for these mapping-specific properties when assigning computations to execution cycles and determining feasible operation chaining under data dependencies and the target clock period.
Together, these choices create a broad implementation space spanning expression forms, resource mappings, and schedules. Exploiting it requires coordinating decisions across multiple synthesis stages.

However, conventional frameworks~\cite{xu_hector_2022,amd_inc_vitis_2025-1,nigam_compiler_2021,kim_unifying_2024,ferrandi_invited_2021,canis_legup_2013} constrain its exploration along two dimensions.
In the resource dimension, conventional HLS compilers rely on simplified, compiler-internal resource models. Vitis HLS~\cite{amd_inc_vitis_2025-1} exposes per-operation implementation and latency choices through \texttt{BIND\_OP}, whereas its automatic DSP fusion recognizes only AMA-like expression patterns encoded in the compiler. Other supported patterns remain unavailable to automatic synthesis without compiler extensions~\cite{smith_fpga_2024, smith_scaling_2024}.
Furthermore, transformation, scheduling, and mapping are performed sequentially. Transformation first commits to one expression without evaluating how equivalent forms would perform under different mappings and schedules, leaving later phases to operate on the selected algebraic structure. Scheduling then assigns execution cycles using coarse mapping estimates, which may prevent feasible operation chaining or assign operations to different cycles even when they could be mapped together onto the same resource. Mapping considers only choices consistent with earlier decisions.
Consequently, many combinations of algebraic forms, resource mappings, and schedules are pruned before their end-to-end performance can be evaluated, leaving much of the target's hardware heterogeneity unexplored.

Prior work offers only partial remedies for this problem.
Retiming~\cite{bryant_optimizing_1983} relocates registers in an existing datapath to improve the clock period.
Feedback-guided scheduling~\cite{ye_subgraph_2024,zheng_fast_2014} uses downstream synthesis results to revise cycle boundaries with more accurate timing information.
Both revisit timing only after the computation graph has been chosen, so earlier algebraic alternatives remain unavailable.
~\cite{tan_mapping-aware_2015,wang_mapbuf_2023} couple scheduling with LUT technology mapping or buffer insertion.
Yet their problem formulations restrict this coupling to LUT-based circuits and therefore do not capture the functional and timing heterogeneity of FPGA resources.

An e-graph efficiently represents a large set of expressions and the equivalences among them. Equality Saturation (EqSat)~\cite{max_willsey_egg_2021,zhang_better_2023} applies rewrite rules over this representation and defers choosing a particular expression until extraction. EqSat has recently been applied to domain-specific compiler optimization~\cite{park_trinity_2026, vanhattum_vectorization_2021,thomas_automatic_2026, briggs_implementation_2024}, datapath optimization~\cite{cheng_seer_2024,coward_automatic_2022,coward_rover_2024, saiki_target-aware_2025} and logic synthesis~\cite{pi_esfo_2023,chen_e-syn_2024,chen_e-morphic_2025,smith_scaling_2024,smith_fpga_2024,hofmann_eqmap_2025}, showing its broad optimization potential.
Our key insight is to use EqSat to preserve hardware heterogeneity alongside algebraic equivalence and expose both to end-to-end scheduling.

Based on this insight, we propose \skyegg, a framework for heterogeneity-aware hardware synthesis via equality saturation.
\skyegg unifies algebraic transformations and target-specific hardware mappings by expressing both as rewrite rules over the same e-graph.
Applying both rule sets exposes algebraically equivalent expression forms to target mappings, while each matched mapping retains its resource configuration, pipeline latency, and other characterized timing properties.
\skyegg formulates the transformation- and mapping-aware scheduling problem over the saturated e-graph as an Integer Linear Programming (ILP) problem.
The ILP unifies transformation, mapping, and scheduling under one end-to-end objective by jointly selecting an equivalent expression, its hardware mappings, and execution cycles to minimize latency under the target frequency and hardware timing constraints.
To improve scalability, we introduce a heuristic that explores the same heterogeneous space efficiently.
\skyegg is open source.

This work makes the following contributions:

\begin{itemize}[leftmargin=*]
\item We present \skyegg, an open-source heterogeneity-aware hardware synthesis framework for end-to-end exploration of the heterogeneous FPGA implementation space.

\item We encode resource capabilities and configurations as extensible mapping rules, unifying resource heterogeneity and algebraic equivalence within equality saturation.

\item We propose a transformation- and mapping-aware scheduler that jointly selects an equivalent form, its mappings, and execution cycles to minimize end-to-end latency via an exact ILP or scalable heuristic.
\end{itemize}

\boldparagraph{Evaluation.}
We evaluate \skyegg across multiple target frequencies on two FPGA platforms, comparing it with two state-of-the-art commercial HLS tools and an extended MapBuf implementation with operator-level DSP mapping support. \skyegg achieves average speedups of 3.12\x over Vitis HLS, 3.18\x over Altera HLS, and 1.37\x over MapBuf, while all evaluated designs meet timing. It maintains FF usage moderate while effectively utilizing LUTs and DSPs to achieve performance gains. The heuristic solver handles designs with up to 600 operations within 3 seconds, stays within 10\% of exact ILP latency on common test cases, and provides warm starts that extend ILP scalability beyond 400 operations.

\section{Background and Motivation}

We first walk through heterogeneous FPGA resources and conventional synthesis flows, then present the background on Equality Saturation and examples that motivate our work.

\subsection{Hardware Synthesis Targeting Heterogeneous FPGA Resources}
\label{sec:hardware-synthesis}

\begin{figure}[t]
  \centering
  \includegraphics[width=\linewidth]{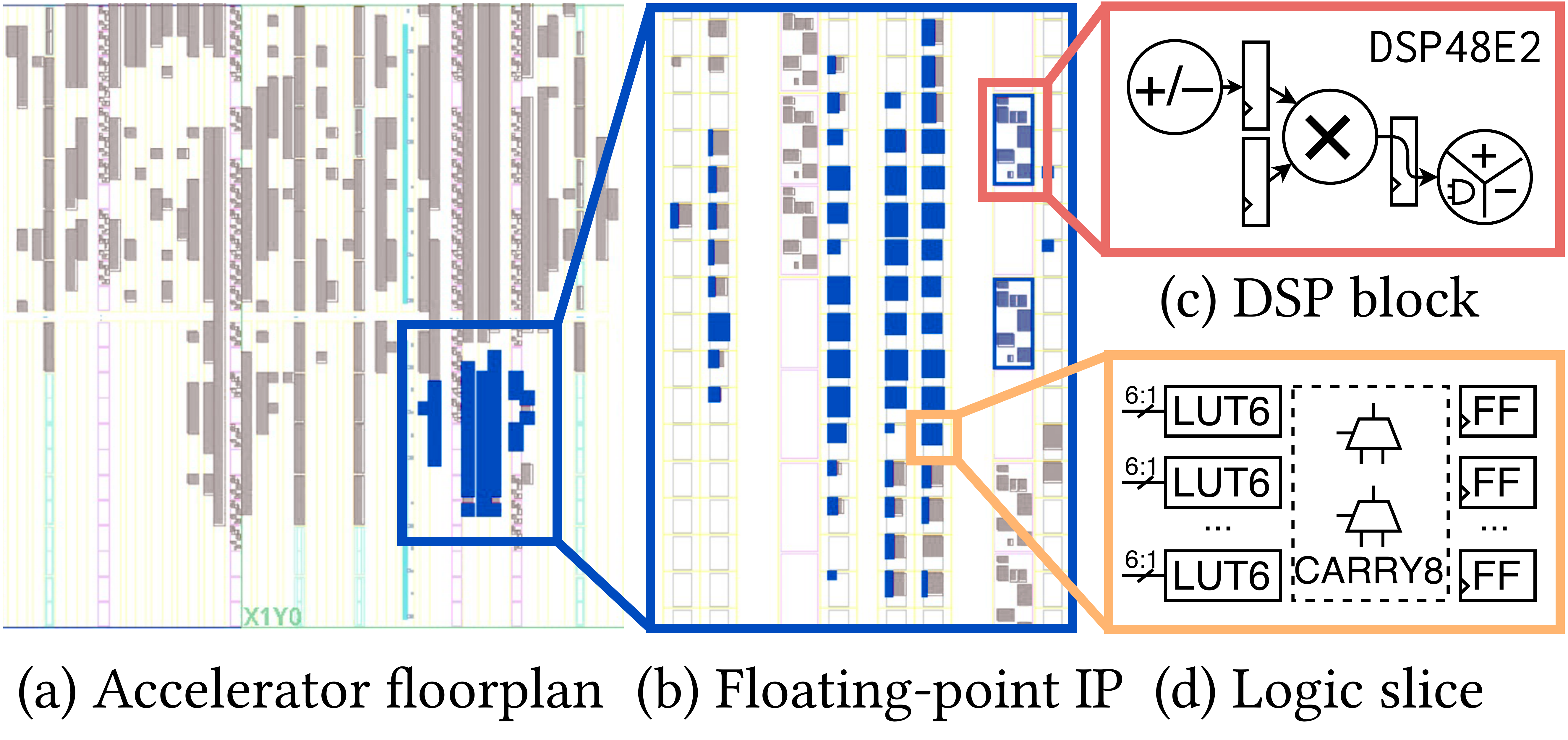}
  \caption{Heterogeneous computing resources in a modern FPGA accelerator.}
  \label{fig:hetero}
\end{figure}

Modern FPGAs are heterogeneous, combining flexible programmable logic with specialized and configurable arithmetic resources. The accelerator floorplan in \autoref{fig:hetero} illustrates how these resource types coexist within a single design. LUTs, FFs, and carry chains provide a general substrate for arbitrary logic, while Digital Signal Processors (DSPs) offer dedicated high-speed arithmetic datapaths.
Xilinx's DSP48E2~\cite{amd_inc_ultrascale_2021} combines a pre-adder, multiplier, and ALU to implement fused expressions such as \texttt{(A+D)*B+C} in one block. Its five pipeline stages can be independently enabled or bypassed, trading cycle latency for maximum frequencies ranging from roughly 304MHz to 600MHz~\cite{amd_inc_kintex_2025}. These datapath and pipeline choices expose hundreds of configuration parameters. Altera DSP blocks~\cite{intel_agilex_2026} support both fixed-point and floating-point arithmetic, and provide split-multiplier modes for efficient low-precision computation. Parameterized dividers and floating-point IPs further expand this space through latency and resource presets that trade DSP, RAM, and LUT usage across different internal algorithms~\cite{xilinx_inc_floating-point_2020}. Selecting among these resources requires balancing latency, resource use, and frequency, while conservative configurations can leave performance suboptimal.

Conventional hardware synthesis frameworks~\cite{xu_hector_2022,amd_inc_vitis_2025-1,canis_legup_2013} follow three sequential steps. First, transformation applies canonicalization and algebraic rewriting to make the operation graph more amenable to hardware implementation, thereby determining the arithmetic form presented to downstream synthesis. Second, scheduling assigns its operations to execution cycles under data dependencies and the target clock period, determining their temporal overlap and register boundaries. Finally, mapping binds each scheduled operation to a target-specific resource type and configuration from the target library before lowering the design toward RTL. Consequently, fixed arithmetic forms and schedules limit how heterogeneous resources can be used.

\subsection{Equality Saturation}

\begin{figure}[t]
  \centering
  \includegraphics[width=0.8\linewidth]{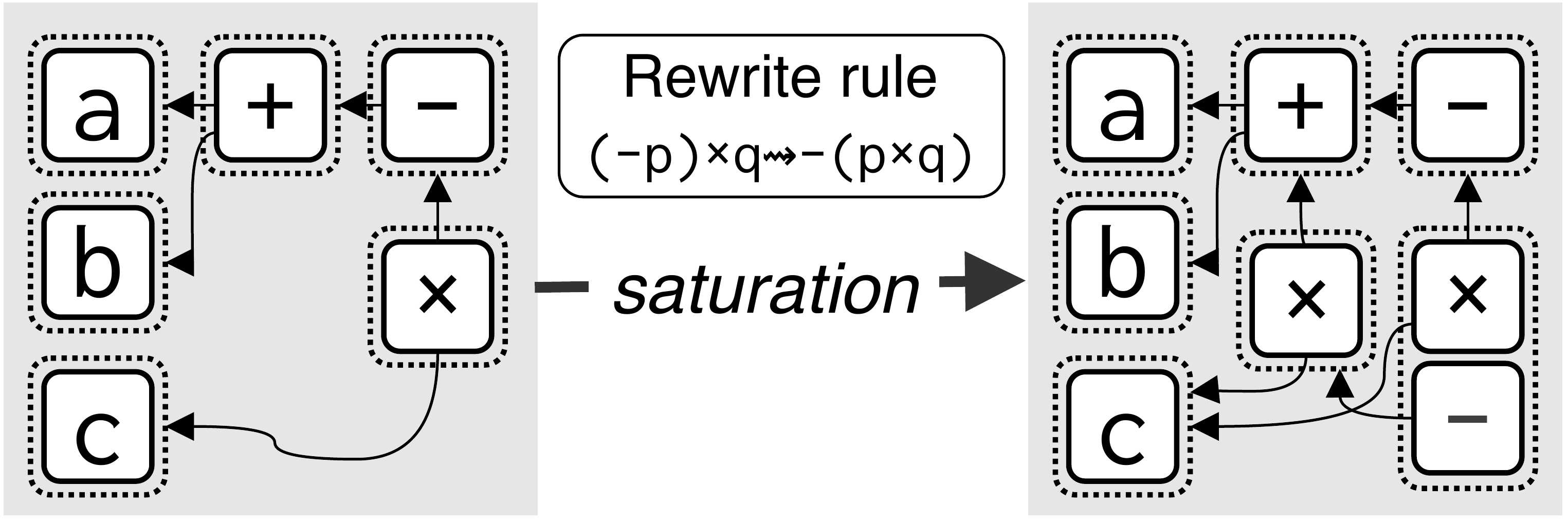}
  \caption{An e-graph groups equivalent terms into e-classes, and saturation adds alternatives from rewrite rules.}
  \label{fig:egraph-background}
\end{figure}

Equality saturation (EqSat)~\cite{max_willsey_egg_2021,zhang_better_2023} non-destructively explores equivalent forms of a program. An e-graph represents equivalent \emph{terms} (or \emph{expressions}) as equivalence classes. Each equivalence class, called an \emph{e-class}, contains a group of \emph{e-nodes}, and each e-node represents a function symbol (also called a \emph{constructor}) applied to argument e-classes. This allows structurally different terms share common subexpressions. EqSat exploration proceeds over this e-graph in two stages, \emph{saturation} and \emph{extraction}. During saturation, equivalent rewrite rules are repeatedly applied until no new equivalences can be discovered or a time budget is reached. When a rule matches, its right-hand side is added to the e-graph and its root e-class is unioned with that of the matched left-hand side, preserving both forms. As shown in \autoref{fig:egraph-background}, applying

$(-p)\!\times\!q \mathrel{\mkern-2mu\rightsquigarrow\mkern-2mu} -(p\!\times\!q)$
to \texttt{-(a+b)*c} adds \texttt{-((a+b)*c)} as an equivalent form in the root e-class. Extraction then selects best terms from the saturated e-graph according to a cost function. This compact exploration keeps equivalent forms available for evaluation with both target-resource characteristics and downstream synthesis decisions.

\subsection{Motivation}
\label{sec:motivation}

\begin{figure}[t]
\centering
\includegraphics[width=\linewidth]{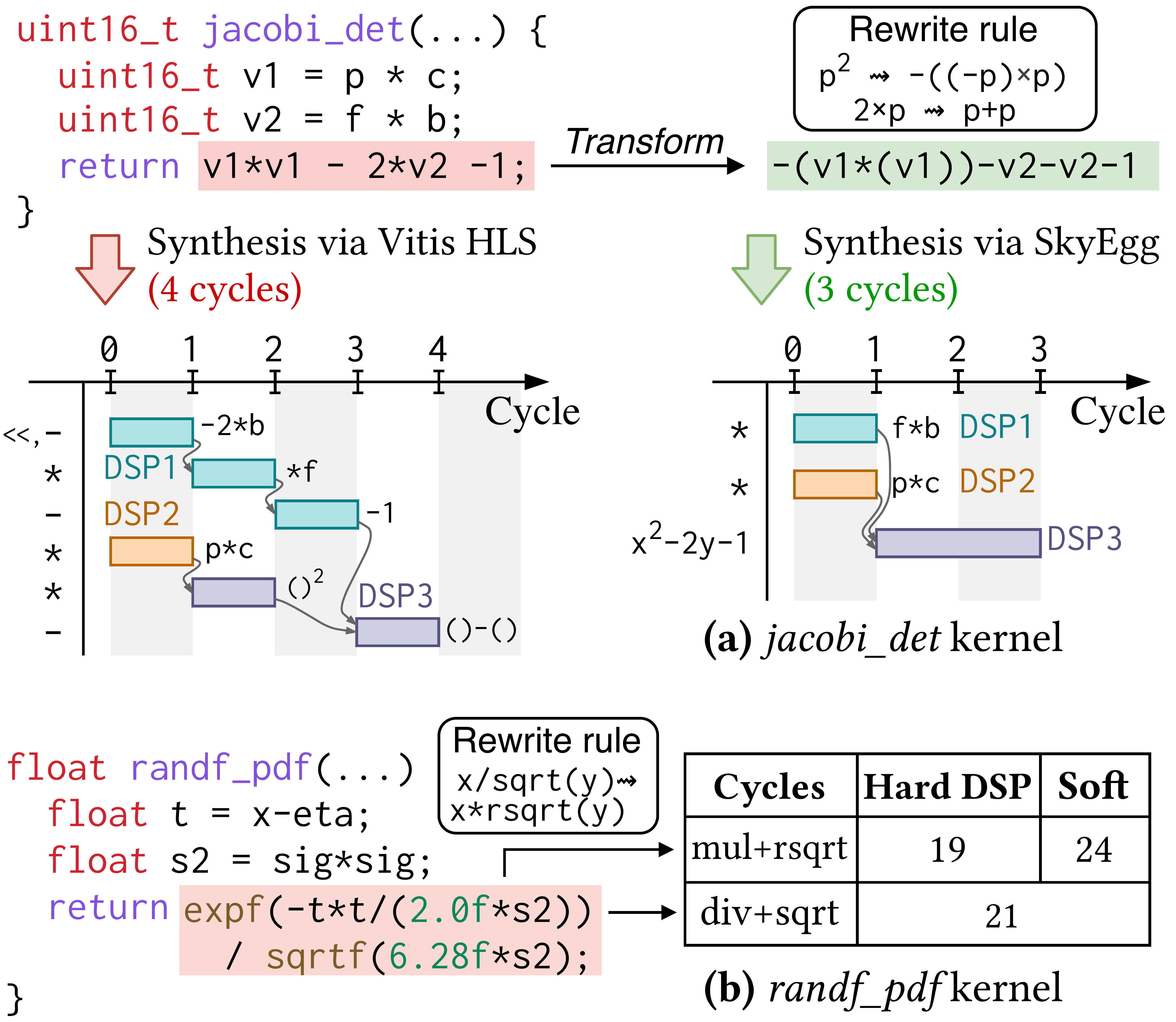}
\caption{Performance losses from missed hardware-dependent opportunities. (a) An equivalent form enables a lower-latency fused DSP mapping. (b) The same rewrite can improve or degrade latency with different mappings.}
\label{fig:motivate}
\end{figure}

We show two representative cases synthesized by Vitis HLS 2024.1, targeting Xilinx XCKU3P FPGA at 400MHz.

\boldparagraph{Hidden Fused Mapping Opportunities.} The first example is a fixed-point Jacobian determinant, shown in \autoref{fig:motivate}(a). Vitis HLS schedules the original expression in four cycles, computing \texttt{-2*b} and \texttt{p*c} in parallel, then \texttt{(-2*b)*f} and \texttt{(p*c)}\textsuperscript{2}, before incorporating \texttt{-1} and performing the final subtraction. A technology mapper then fuses these operations into three DSPs. Yet a lower-latency fused mapping requires the form \texttt{-(x*(-x))-y-y-1}, which can be derived from \texttt{x}\textsuperscript{2}\texttt{-2*y-1} by applying the two rewrite rules shown in \autoref{fig:motivate}(a). This form allows \skyegg to combine the DSP's pre-adder, multiplier, and ALU in a single datapath\footnote{Using ALU mode \texttt{4'b0010}, where P = \textasciitilde(W + X:Y + Z)}. Setting both DSP2 and DSP3 to two-cycle latency reshapes the schedule and reduces end-to-end latency by 25\%.
Here, algebraic reasoning alone provides no basis for seeking this particular form, whose value arises only from this particular DSP mapping.

\boldparagraph{Rewrite Selection without Mapping Context.} Discovering such forms is only part of the problem. Vitis HLS synthesizes the FP32 kernel \texttt{randnf\_pdf} in \autoref{fig:motivate}(b) into a 115-cycle design using 13 DSPs. For \texttt{x/sqrt(y)}, it instantiates a divider and a square-root IP with 29 cycles of latency each, although our profiling shows that both IPs can run at over 800MHz at this setting. This conservative configuration introduces unnecessary cycles, tuning their latency alone reduces the total to 21 cycles. Another equivalent form, \texttt{x*rsqrt(y)} for \texttt{x/sqrt(y)}, replaces the divider and square root with a multiplier and reciprocal square root. With setting "\textit{Full DSP Usage}", the rewritten form takes 19 cycles, two fewer than the tuned original. With DSPs disabled, however, it takes 24 cycles and falls behind. The rewrite's latency impact therefore depends on the available mappings and cannot be determined by a transformation-only cost.

\skyegg addresses both problems by retaining equivalent forms and target mappings in the same e-graph, allowing hardware mappings to recognize useful forms. It then unifies expression-form selection, hardware mapping, and scheduling as a joint optimization over the saturated e-graph, selecting the lower-latency form under the available mappings and timing constraints. Together, these capabilities turn hardware-dependent algebraic opportunities into end-to-end performance gains.

\begin{figure}[t]
  \centering
  \includegraphics[width=.7\linewidth]{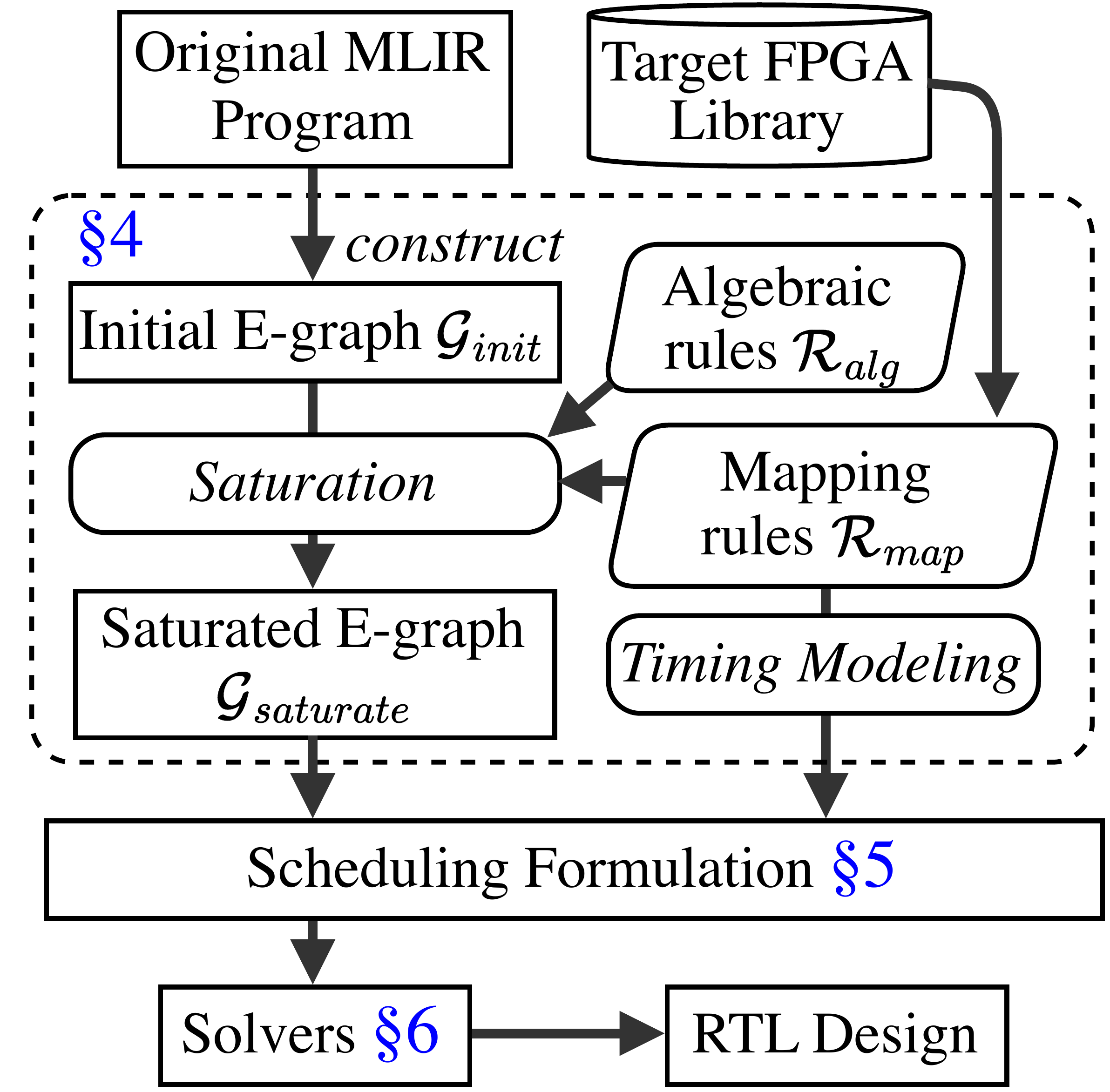}
  \caption{The end-to-end workflow of \skyegg.}
  \label{fig:overview}
\end{figure}

\section{Overview}
\label{sec:overview}

\autoref{fig:overview} presents the workflow of \skyegg. \skyegg takes an MLIR~\cite{lattner_mlir_2020} program and a target FPGA library as inputs, and constructs the initial e-graph ($\mathcal{G}_\texttt{init}$) from the operations in the MLIR. From the target library, \skyegg loads mapping rewrite rules ($\mathcal{R}_{map}$) and characterized timing properties of the hardware realizations. Each mapping rule connects a software expression matched by its \emph{matcher} pattern to an equivalent target-specific hardware mapping (\autoref{sec:egraph}).

During saturation, \skyegg unifies algebraic transformation and target-specific mapping through a shared rewrite interface, applying the algebraic rules ($\mathcal{R}_{alg}$) and mapping rules ($\mathcal{R}_{map}$) to the same e-graph to produce $\mathcal{G}_\texttt{saturate}$. The resulting saturated e-graph preserves hardware heterogeneity by allowing target mappings to match algebraically equivalent expression forms while retaining the resource configuration and timing properties of each matched realization. \skyegg then replaces the extraction phase in EqSat with transformation- and mapping-aware scheduling that selects an equivalent expression and its hardware mappings from the saturated e-graph while assigning execution cycles (\autoref{sec:formulation}). These coupled decisions are solved either exactly through an ILP or approximately with a scalable heuristic (\autoref{sec:solving}), after which \skyegg generates synthesizable SystemVerilog from the selected mappings and schedule.

\section{E-graph Representation and Modeling}
\label{sec:egraph}

\begin{figure}[t]
  \centering
  \includegraphics[width=\linewidth]{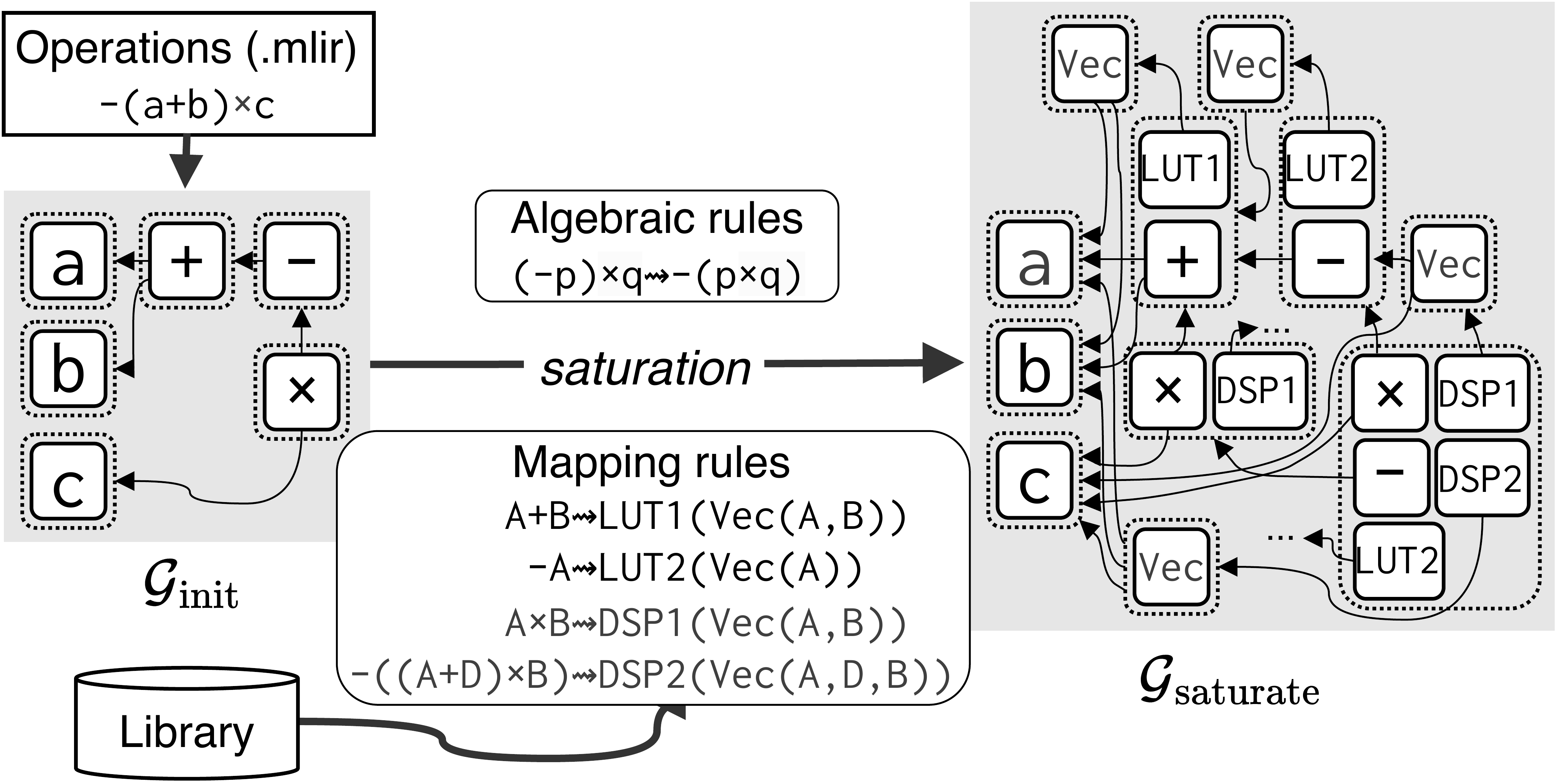}
  \caption{E-graph with algebraic and mapping rules jointly applied during saturation.}
  \label{fig:egraph}
\end{figure}

As the basis for representing the heterogeneous implementation space, \skyegg first constructs an e-graph from the input MLIR expressions. Each SSA value becomes an e-node whose constructor is its defining MLIR operation and whose arguments reference the operand e-classes. For example, the MLIR program for \texttt{-(a+b)*c} produces $\mathcal{G}_{\texttt{init}}$ in \autoref{fig:egraph}.

\subsection{Mapping Representation in E-graph}

The initially created e-graph does not provide mapping information for the contained operations. Representing the design space of possible mappings requires encoding mapping candidates within the e-graph data structure. We define a specific class of e-nodes to represent mapping candidates, called \emph{mapping e-nodes}: \texttt{
id(Vec(arg1, arg2, ...))}. The \texttt{id} uniquely identifies a mapping, while the \texttt{Vec} constructor lists its inputs in hardware port order.

An FPGA resource may provide one or more configurations for the same computation pattern. Simple LUT-based mappings need no configuration choices, while configurable resources such as DSP blocks requires encoding their pipeline depths and timing properties alongside datapath selections. \skyegg represents each configuration as a distinct mapping e-node, with its parameters encoded in the \texttt{id}. Each mapping e-node shares an e-class with functionally equivalent e-nodes and corresponds to a fixed circuit that can be directly instantiated during code generation. Hereafter, the term \emph{mapping} refers to a specific mapping-configuration pair with deterministic scheduling properties.

\boldparagraph{Mapping Rules.}
We formulate the relationship between a computation pattern and a hardware mapping as a \emph{mapping rule}. During saturation, each \emph{mapping rule} matches expressions in the e-graph and inserts the corresponding \emph{mapping e-node} into the root e-class of each match. Specifically, we define each mapping rule as three parts:
\begin{enumerate}[leftmargin=*]
  \item \emph{Matcher}: an algebraic pattern that matches one or a group of operations. It represents the software functionality of the mapping, describing what computation it can perform.
  \item \emph{Applier}: a mapping e-node pattern with the variables from the \emph{matcher} as the input arguments.
  \item \emph{Condition}: a set of constraints on the mapping arguments. It generally contains data type or bitwidth requirements.
\end{enumerate}

For each computation pattern supported by an FPGA resource, \skyegg generates one mapping rule per configuration. A pattern without configuration choices produces one rule pairing one matcher with one applier. Multiple configurations of the same computation pattern share the matcher but use distinct appliers, which insert mapping e-nodes with their respective resource and timing properties.

For example, \autoref{fig:egraph} shows four \emph{mapping rules} that are provided by the target implementation library. Each rule is described in the form: $\textit{matcher} \rightsquigarrow \textit{applier}$, with \emph{conditions} omitted for brevity. The figure also omits the timing characteristics associated with each mapping. The first two rules map addition and negation to \texttt{LUT1} and \texttt{LUT2}, respectively. The last two map multiplication and a fused arithmetic pattern to \texttt{DSP1} and \texttt{DSP2}, with applier arguments ordered by the DSP input ports. The two DSP rules require integer inputs satisfying the DSP48E2 port constraints~\cite{amd_inc_ultrascale_2021}. Using this representation, \skyegg uniformly encodes mappings for heterogeneous FPGA resources. New mappings can therefore be added through corresponding rules without modifying the mapping algorithm or the scheduling formulation.

\boldparagraph{Saturation.} \skyegg constructs the saturated e-graph by iteratively applying mapping rewrite rules ($\mathcal{R}_{\texttt{map}}$) alongside algebraic rewrite rules ($\mathcal{R}_{\texttt{alg}}$). The algebraic rules expose equivalent forms, allowing mapping rules to match computation patterns obscured in the original expression. \skyegg can incorporate compatible algebraic rewrite rules from prior e-graph-based optimizations~\cite{cheng_seer_2024,coward_automatic_2022}.
To keep the saturated graph acyclic for downstream timing analysis and scheduling, \skyegg prunes algebraic rewrite rules that may introduce cyclic e-class dependencies, especially rules that expand a single term, such as $x \rightsquigarrow x+0$.

\autoref{fig:egraph} illustrates how the two rule sets interact. For the original expression, the addition, negation, and multiplication match \texttt{LUT1}, \texttt{LUT2}, and \texttt{DSP1}, respectively. The algebraic rule $(-p)\times q \rightsquigarrow -(p\times q)$ adds the form \texttt{-((a+b)*c)}, represented by the \emph{negation} (\texttt{-}) e-node in the root e-class on the lower left. This form matches the fused arithmetic pattern of the \texttt{DSP2} rule, which inserts a fused DSP mapping into the root e-class. The saturated e-graph retains both the mappings for the original expression and the fused DSP mapping. Together, the two rule sets allow target mappings to be explored across algebraically equivalent forms.

\subsection{Timing Analysis over E-graph}
\label{sec:timing-properties}

\newcommand{\tin}{$t_\text{in}$\xspace}
\newcommand{\tinport}[1]{{$t_\text{in,#1}$}\xspace}
\newcommand{\tinid}[1]{$t^\text{#1}_\text{in}$\xspace}
\newcommand{\tinidport}[2]{{$t^\text{#1}_\text{in,#2}$}\xspace}
\newcommand{\tout}{$t_\text{out}$\xspace}
\newcommand{\tcycle}{$t_\text{rr}$\xspace}
\newcommand{\tnet}{$t_\text{net}$\xspace}
\newcommand{\tclkq}{$t_\text{cq}$\xspace}
\newcommand{\tsu}{$t_\text{su}$\xspace}

The saturated e-graph contains different mappings with configurations that expose different timing characteristics. \skyegg analyzes their structural paths and corresponding delays directly on the e-graph to provide the timing information required by scheduling.

\begin{figure}[t]
  \centering
  \includegraphics[width=\linewidth]{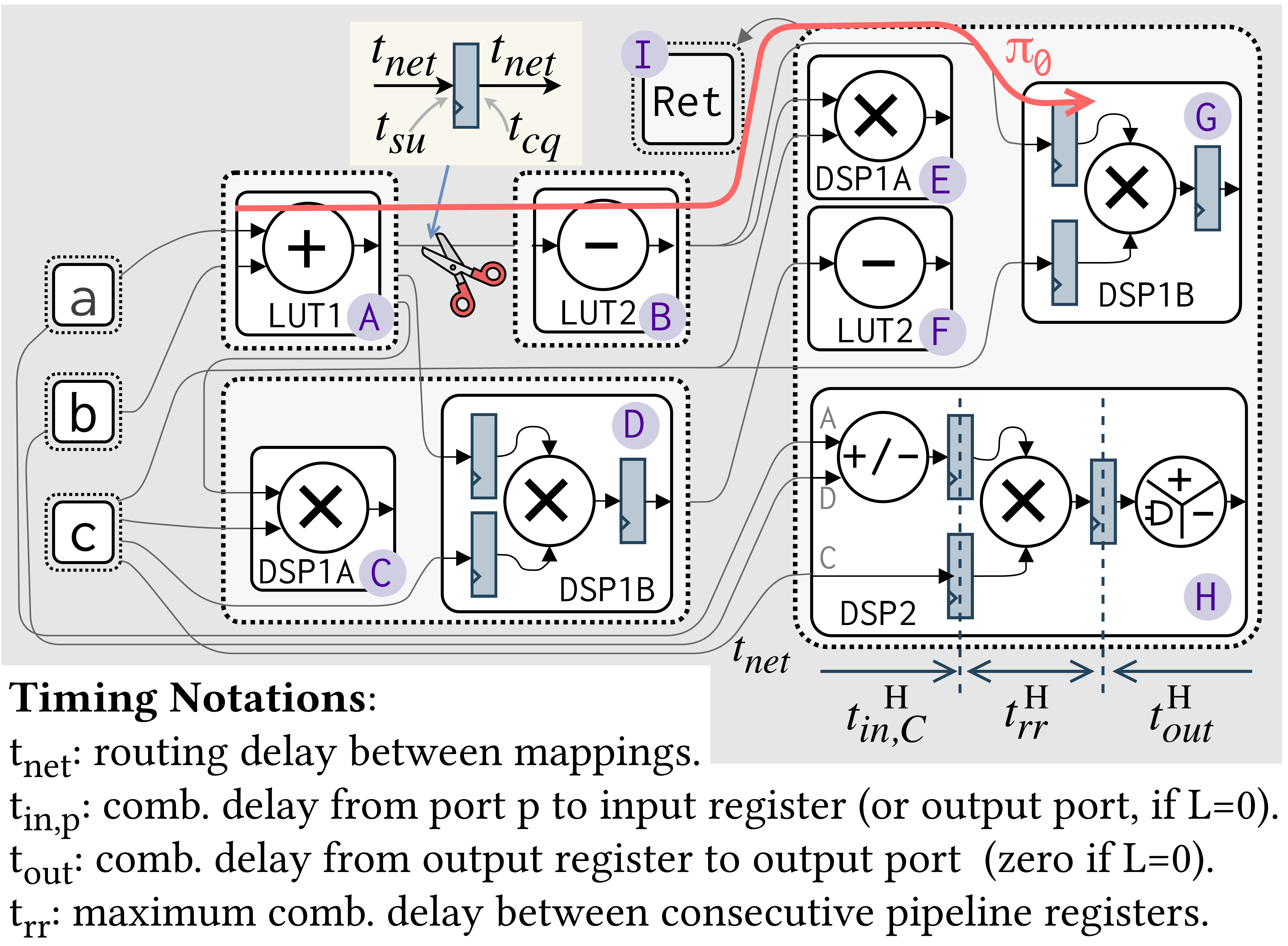}
  \caption{Timing analysis over a saturated e-graph and associated timing notation.}
  \label{fig:timing-model-example}
\end{figure}

\boldparagraph{Timing Properties of Mappings.} We establish a multiple-input-single-output model to capture each target mapping's timing properties. We continue to use the saturated \texttt{-(a+b)*c} example in \autoref{fig:timing-model-example}, where we removed the software e-node and expand the timing model of each \textit{mapping e-node}. The \texttt{Vec} e-nodes are omitted from the figure for brevity. Alongside the latency ($L$) of the mapping, we define a set of timing notations, labeled on the bottom-right of the figure.

To obtain this timing model for each mapping, we develop an automated offline profiler based on the vendor's FPGA synthesis and fitting tools. The profiler instantiates the mapping, applies timing constraints, runs synthesis and fitting, and extracts these metrics from the resulting reports. This post-fitting timing provides a good approximation, sufficient to guide further optimization. The extracted properties are recorded in the mapping library and reused across input programs over the same FPGA platform. Further analysis is automatically performed by \skyegg using these models.

\boldparagraph{Timing Paths and Register Cuts.} The register-to-register delay \tcycle obtained by the profiler captures only the delay between internal pipeline registers. We further define the \emph{path delay} to capture the combinational delay along a specific path connecting two mappings. The \emph{path delay} of a valid structural path $\pi$ connecting mapping $m_i$ to mapping $m_j$, denoted as $t_{\text{path}}(\pi)$, is defined as the accumulated combinational delay along that path. A valid path may traverse multiple combinational intermediate mappings, and its path delay can be calculated by summing the output delay (\tout) of the source mapping $m_i$ (or \tin if $L_{m_i}=0$), the input delays (\tin) of other mappings it passed, and the routing delays ($t_{\text{net}}$) between them. In \autoref{fig:timing-model-example}, the path delay of the labeled path $\pi_0$ is therefore the sum of \tinidport{A}{A}, \tinid{B}, \tinidport{G}{A}, and two \tnet.

If $t_{\text{path}}(\pi_0)$ exceeds timing constraints, a pipeline register must be inserted to break the combinational chain. As shown in the yellow box, the net effect is to split the original path into two stages, and form two new timing paths on each side of the register. The path delay of the left side now becomes the sum of \tinidport{A}{A}, \tnet, and a setup delay \tsu, and the delay of the right side is sum of \tclkq, \tinid{B}, \tinidport{G}{A}, and two \tnet. By inserting this register, we trade one additional cycle of latency for meeting timing requirements, while also introduce additional data path delays for both the upstream and downstream portions of the pipeline, which requires careful handling.

\boldparagraph{Timing Analysis of Mapping Alternatives.} In conventional data-flow graphs (DFGs), the operators and port connections along a path are fixed. However, in an e-graph, an e-class may contain multiple equivalent mapping e-nodes, creating paths with different delays and mapping choices between the same endpoint. Timing analysis must enumerate these alternatives because scheduling may select any of them. We therefore use \textit{path group}, $\Pi(m_i,m_j)$, to denote the set of valid combinational paths from mapping $m_i$ to $m_j$.

\newcommand{\sto}{\!\mathrel{\scalebox{0.6}[1]{$\to$}}\!}

Take path group $\Pi(m_A,m_I)$ in \autoref{fig:timing-model-example} as an example, it contains path $(m_A\sto m_B\sto m_E\sto m_I)$ and $(m_A\sto m_C\sto m_F\sto m_I)$, with delays derived from the timing model above. Starting from the predecessor e-class of $m_I$, $\{m_E, m_F, m_G, m_H\}$, if the final extraction selects $m_E$, $m_B$ must also be selected, requiring the former path to either satisfy the constraint or be cut. If $m_F$ is selected instead, $m_C$ must also be selected, and the latter path must then be considered. Selecting $m_G$ or $m_H$ leaves at least one mapping e-node in each path unselected, so neither path needs to be accounted for. \skyegg enumerates these alternative paths with their delays and provides them to downstream scheduling, which enforces the timing constraints for the selected implementation.

\boldparagraph{Top-k Path Delays.} Mapping alternatives across successive e-classes can produce a rapidly growing number of structural paths, making it impractical to materialize the complete $\Pi(m_i,m_j)$ in large designs. To bound this analysis, \skyegg summarizes each mapping pair by its $k$ longest path delays, preserving multiple high-delay alternatives without propagating the complete path group. The top-$k$ path delays from mapping $m_i$ to mapping $m_j$, denoted as $\mathcal{T}_{\text{top-k}}(m_i,m_j)$, are
\begin{align}
\mathcal{T}_{\text{top-k}}(m_i, m_j)
=
\text{top-k}
\left\{
t_{\text{path}}(\pi)
:
\pi \in \Pi(m_i, m_j)
\right\}.
\label{def:top-k-path-delays}
\end{align}

The top-$k$ path delays can be calculated recursively over the acyclic e-graph as follows:
{\small
\thinmuskip=1mu
\medmuskip=1mu
\thickmuskip=2mu
\setlength{\jot}{1pt}
\begin{align}
\mathcal{T}_{\text{top-k}}(m_i,m_j)
&\!=\!\operatorname{top}_k\!\bigl(
\bigcup_{\substack{(ec_l,p)\in\\\operatorname{pred}(m_j)}}
\mkern-14mu\bigl\{d+t_{\mathrm{net}}+t^{m_j}_{\mathrm{in},p}
\mid d\in\mathcal{T}_{\text{top-k}}(m_i,ec_l)\bigr\}
\bigr),
\label{eq:top-k-mapping-recursion}\\
\mathcal{T}_{\text{top-k}}(m_i,ec_j)
&\!=\!\operatorname{top}_k\!\bigl(
\bigcup_{m_j\in ec_j}
\mathcal{T}_{\text{top-k}}(m_i,m_j)
\bigr).
\label{eq:top-k-eclass-recursion}
\end{align}
}
\autoref{eq:top-k-mapping-recursion} extends partial delays through mapping $m_j$, while \autoref{eq:top-k-eclass-recursion} merges the mapping alternatives in e-class $ec_j$. The recursion follows the path boundaries defined above, and each retained delay remains associated with its structural path for the corresponding chaining constraint.

With these timing analysis techniques, \skyegg unifies algebraic equivalences, heterogeneous mapping and configuration alternatives, and their timing implications within a single e-graph representation, providing the design-space information required by further optimization formulation.

\section{Problem Formulation}
\label{sec:formulation}

\begin{table}[t]
  \caption{Notations for problem formulation}
  \small
  \renewcommand{\arraystretch}{0.95}
  \label{tab:notations}
  \begin{tabular}{ll}
    \toprule
    Symbol & Description \\
    \midrule
    $\{m_i\}_{i=0}^{N_\text{node}}$   & set of \emph{mapping} e-nodes in $\mathcal{G}_{\texttt{saturate}}$ \\
    $\{cls_j\}_{j=0}^{N_\text{class}}$ & set of \emph{e-classes} in $\mathcal{G}_{\texttt{saturate}}$ \\
    $cls_{root}$        & root e-class in $\mathcal{G}_{\texttt{saturate}}$ \\
    $m_k \in cls_j$     & \emph{e-class} $cls_j$ contains the e-node $m_k$ \\
    $child_{m_i}$       & set of \emph{child} e-classes of the e-node $m_i$ \\
    $f_{cls_j}$         & finish time of the e-class $cls_j$ \\
    $b_{cls_j}$         & whether the e-class $cls_j$ is selected \\
    $s_{m_i}$/ $f_{m_i}$ & start/finish time of the e-node $m_i$ \\
    $b_{m_i}$           & whether the e-node $m_i$ is selected \\
    \bottomrule
  \end{tabular}
\end{table}

In this section, we formulate the transformation- and mapping-aware scheduling problem. The solution to the problem assigns each software operation to a mapping, allowing one hardware unit to finish multiple operations according to its functionality, and determines the execution clock cycles for each mapping. It's worth noting that only \emph{mapping} e-nodes are considered in the formulation, while other software operation e-nodes are not included.
In addition to the timing notations defined in \autoref{sec:timing-properties}, \autoref{tab:notations} summarizes the notation for the saturated e-graph structure, selection decisions, and scheduling times used in the formulation. Using these notations, we formulate the ILP as follows:

{\small\setlength{\jot}{1pt}
\begin{subequations}
  \label{eq:formulation}
  \begin{align}
    & \text{minimize}_{f_{cls_j}, b_{cls_j}, s_{m_i}, f_{m_i}, b_{m_i}}\:\;f_{cls_{root}} + \alpha \sum_{i}^{N} b_{m_i} \\
    \text{s.t.} \quad & b_{cls_{root}} = 1 \label{eq:root-selected}\\
    & \forall j, b_{cls_j} \leq \sum_{m_i \in cls_j} b_{m_i} \label{eq:class-complete}\\
    & \forall i, \forall cls_j \in child_{m_i}, b_{m_i} \leq b_{cls_j} \label{eq:node-complete}\\
    & \forall i, \forall cls_j \in child_{m_i}, s_{m_i} \geq f_{cls_j} \label{eq:node-start}\\
    & \forall i, f_{m_i} = s_{m_i} + L_{m_i} \label{eq:node-finish}\\
    & \forall j, \forall m_i \in cls_j, b_{m_i}=1 \Rightarrow f_{cls_j} \geq f_{m_i} \label{eq:class-finish}\\
    & \forall \pi \in \Pi(m_i, m_j), \bigl(\bigwedge_{m \in \pi}(b_m=1)\bigr) \Rightarrow s_{m_j} \geq f_{m_i} + \text{cuts}(t_\text{path}(\pi)) \label{eq:chaining-constraint}\\
    & \forall m_i, b_{m_i} \in \{0, 1\} \\
    & \forall cls_j, b_{cls_j} \in \{0, 1\}
  \end{align}
\end{subequations}
}
where $\Rightarrow$ denotes an indicator constraint~\cite{bonami_mathematical_2015} that enforces the constraint on its right when its left-hand-side holds.

In the formulation, the objective function is to minimize the total latency of the program, represented as the finish time of the root e-class, $f_{cls_{root}}$. We use a penalty term with a small coefficient $\alpha$ to reduce the number of selected mappings for achieving the best possible latency. Among solutions with the same minimum latency, this penalty favors realizing the computation with fewer mappings. Such solutions typically consolidate more computation into each selected resource, better exploiting its capabilities while avoiding unnecessary resource use. The formulation includes two classes of constraints, as described below.

\boldparagraph{Completeness Constraints:} ensure the correctness of mapping to provide the same functionality as the original program. \autoref{eq:root-selected} ensures the root e-class is selected. \autoref{eq:class-complete} ensures when an e-class is selected, at least one of its \emph{mapping} e-nodes is selected, representing that the e-class's value is properly computed by the selected mapping. \autoref{eq:node-complete} ensures that when a \emph{mapping} e-node is selected, its children e-classes are also selected, representing that the argument values of the mapping are provided.

\boldparagraph{Scheduling Constraints:} require legal scheduling of the mappings. \autoref{eq:node-start} defines the \emph{dependency constraint} of scheduling, ensuring the children e-classes generate the argument values before the mapping e-node starts. \autoref{eq:node-finish} defines the \emph{latency constraint} to ensure the finish time and start time satisfy the latency property of the mapping. \autoref{eq:class-finish} defines the \emph{selection constraint}, which is special for the scheduling on an e-graph: the finish time of an e-class must not be earlier than the finish time of an included mapping e-node if the e-node is selected. During code generation, each operand is delayed from its e-class finish time to the consuming mapping's common start time, aligning inputs across reconvergent paths.

\emph{Chaining constraints} are also necessary to ensure the operating frequency of the synthesized design meets the target. If all implementation e-nodes on the path $\pi$ from $m_i$ to $m_j$ are selected, as required by the indicator condition, $\text{cuts}(t_\text{path}(\pi))$ \emph{registers} should be inserted to cut the combinational path if the path delay exceeds the clock period $T_{\text{clk}}$. Chaining registers are therefore inserted only between mappings. According to the timing model exemplified by \autoref{fig:timing-model-example}, we have the following constraint for timing closure:
\begin{equation}
  t_{\text{path}}(\pi) + q(t_\text{su} + t_{\text{cq}} + t_{\text{net}}) \leq (q+1)T_{\text{clk}} \label{eq:chaining-inequality}
\end{equation}
The constraint states that if we need $q$ registers to cut the combinational path, and the path delay plus the total extra wiring delay, register setup, and clock-to-Q delays should be less than the total delay of $q+1$ split clock cycles. According to \autoref{eq:chaining-inequality}, we calculate the minimum number of registers needed to cut the path:
\begin{equation}
  \text{cuts}(t_{\text{path}}(\pi)) = \left \lceil \frac{t_\text{path}(\pi)-T_{\text{clk}}}{(T_{\text{clk}} - (t_\text{su} + t_{\text{cq}} + t_{\text{net}}))} \right \rceil
\end{equation}

\section{Efficient Solving}
\label{sec:solving}

Although solving the ILP formulation in \autoref{eq:formulation} through an exact solver can produce the optimal mapping and scheduling solution with minimized latency and hardware usage, it is computationally expensive, and HLS tools therefore commonly rely on heuristic algorithms to obtain approximate solutions. In this section, we first introduce a top-$k$ path strategy to reduce the chaining constraints and then present a scalable heuristic solver for this problem.

\boldparagraph{Top-k Chaining Constraints.}
\autoref{eq:chaining-constraint} defines a chaining constraint for every structural path between a pair of mapping e-nodes. Instantiating the formulation with all such paths creates a large constraint set and dominates the solving cost. Traditional scheduling approaches, such as SDC scheduling~\cite{cong_efficient_2006}, commonly consider only the longest timing path between each pair. However, when mapping e-nodes on this path are not selected, another path may become the longest selected path without a corresponding constraint, causing clock-period violations through \emph{over-chaining}. To cover such alternatives while bounding the constraint set, \skyegg incorporates into the ILP only the chaining constraints associated with the \emph{top-k} path delays (defined in \autoref{def:top-k-path-delays}) for each mapping pair. We set $k$ to 3 by default to balance solving speed and synthesis quality. Increasing $k$ retains more alternative paths and better guards against over-chaining, but enlarges the constraint set and increases solving time. Decreasing $k$ reduces the solving cost at the risk of omitting a timing-critical alternative selected by the mapping solution.

\begin{figure}[t]
  \begin{lstlisting}[
    commentstyle=\small\ttfamily\textcolor{gray},
    basicstyle=\small\ttfamily,
    frame=none,
    xleftmargin=2.5em,
    lineskip=-0.5pt,
    numbers=left]
def heuristic_solver($\mathcal{G}_\text{saturate}$): # saturated e-graph
# Iterate over e-classes in topological order
  for $cls_j$ in $\mathcal{G}_\text{saturate}$.eclasses.topo_order():
# Determine the earliest start time for each e-node
    for $m_i \in cls_j$:
      $s_{m_i}$ = max([$f_{cls}$, for $cls \in child_{m_i}$])
# Consider the top-k path delays ending at $m_i$
      for $\pi$ in $\mathcal{T}_{\text{top-k}}(m_i)$:
        if all([$b_{m'}$ for $m'$ in $\pi$.enodes]):
          $s_{m_i}$ = max($s_{m_i}$, $f_{\pi\text{.src}}$ + $\text{cuts}(t_\text{path}(\pi))$)
      $f_{m_i}$ = $s_{m_i}$ + $L_{m_i}$
# Select the earliest-finish implementation
    $m$ = $\argmin_{m_i \in cls_j} f_{m_i}$
    $b_{m}$, $f_{cls_j}$ = 1, $f_{m}$
  return $b_{m_i}$, $s_{m_i}$
  \end{lstlisting}
  \caption{The scalable heuristic solver.}
  \label{fig:heuristic-solver}
\end{figure}

\boldparagraph{Scalable Heuristic Solver.}
Even after reducing the chaining constraints, solving the ILP problem with an exact solver remains NP-hard.
For large-scale designs, HLS tools commonly rely on efficient heuristics to obtain high-quality scheduling results within practical compilation time~\cite{canis_legup_2013, ferrandi_invited_2021, paulin_force-directed_1989, chen_deep-reinforcement-learning-based_2019}. \skyegg further expands the design space with equivalent transformations and alternative hardware mappings. To support larger design spaces, we propose a scalable heuristic solver that incrementally constructs a legal solution without invoking the exact ILP backend. \autoref{fig:heuristic-solver} presents the algorithm. It iterates over the e-classes in topological order and selects the mapping with the earliest finish time for each e-class. For every mapping e-node, it considers both \emph{dependency constraints} and the retained \emph{chaining constraints} associated with \emph{top-k} path delays ending at the e-node to determine the start time. The solver runs in time linear in the e-graph size by making greedy decisions without global search and therefore does not guarantee global optimality. Nevertheless, the heuristic considers the same transformation and mapping alternatives as the reduced ILP and enforces the same dependency and retained chaining constraints. These shared choices and constraints also allow the selected mappings, execution cycles, and latency to directly form a feasible starting solution for exact solving. The heuristic latency further limits the latest finish cycle of each e-class, allowing the exact solver to discard choices that cannot improve it. Together, the feasible solution and latency bound reduce the cost of exact solving without compromising global optimality.

\section{Evaluation}
\label{sec:evaluation}

In this section, we compare \skyegg against two commercial HLS toolchains and one research tool to demonstrate how heterogeneity-aware synthesis improves synthesis quality and to evaluate the method's scalability.

\subsection{Evaluation Setup}

We compare \skyegg with AMD Xilinx Vitis HLS 2024.1 and Altera HLS IP Gen Compiler 2026.1. FPGA synthesis toolchain for each platform is Vivado 2024.1 and Quartus Prime Pro 26.1, targeting XCKU3P and AGF027 FPGA, respectively. We also compare \skyegg with MapBuf~\cite{wang_mapbuf_2023}, which we extend to support high-level DFGs, on DSP mapping tasks targeting Xilinx platforms. All experiments are conducted on a server with two AMD EPYC 7763 CPUs with 1.8 TiB of RAM running Ubuntu 22.04. \skyegg uses CP-SAT solver in OR-Tools 9.14~\cite{google_inc_or-tools_2025} for ILP solving with a 1-hour timeout.

\newcommand{\symadd}{$+$}
\newcommand{\symsub}{$-$}
\newcommand{\symmul}{$\times$}
\newcommand{\symdiv}{$/$}
\newcommand{\symarith}{EA}

\begin{table}[t]
\caption{Benchmarks in \skyegg evaluation. EA is $+-\times\div$.}
\centering
\footnotesize
\renewcommand{\arraystretch}{1.2}
\label{tab:eval-cases}
\begin{tabular}{m{0.07\linewidth}m{0.04\linewidth}m{0.40\linewidth}m{0.15\linewidth}m{0.13\linewidth}}
\toprule
Cat. & DT. & Benchmarks & \# Math Op & Math Ops \\ \midrule
Linalg  \xspace\xspace\xspace\cite{l-n_pouchet_polybenchc_2016} & int & durbin, jacobi-1d, jacobi-2d, bicg16, gemver16, gemm16 & 12,6,10,10, 11,63,177,49 & \symarith \\
DSP\cite{amd_inc_xilinxvitis_libraries_2025} & int & lerp, webp\_enc, rope & 33,44,33 & \symarith, $\ll$, $\&$ \\
Math \xspace\xspace\xspace~\cite{gaeddert_joseph_d_and_others_liquid-dsp_2025} & float & bairstow, invgauss, landenf, lngammaf, randnf\_pdf & 51,9,7,20,10 & \symarith, $e^x$, $\sqrt{x}$, $\ln x$ \\
NN  \xspace\xspace\xspace\cite{georgi_gerganov_ggml_2025} & float & GeLU,GeGLU,SiLU,SwiGLU Softmax,layernorm,rmsnorm & 13,14,4,5, 5,12,7 & \symarith, $e^x$, $\ge$, $\sqrt{x}$ \\
Synth & both & randomly generated & 100$\sim$600 & All above\\
\bottomrule
\end{tabular}
\end{table}

\begin{figure*}[t]
  \centering
  \includegraphics[width=\linewidth]{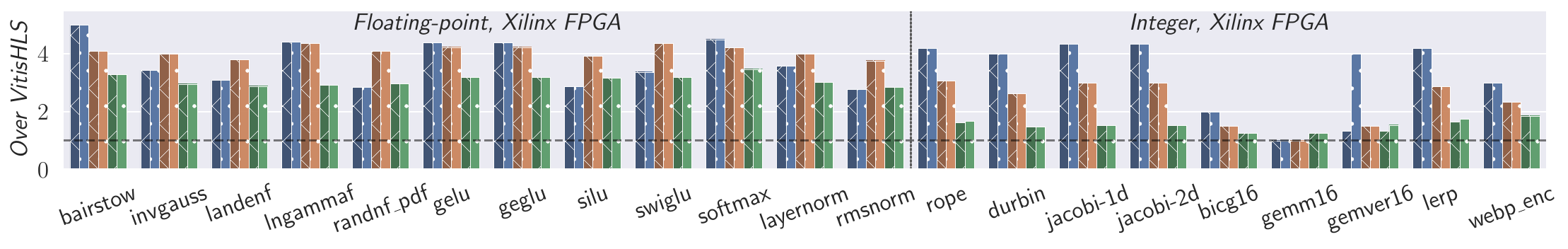}
  \includegraphics[width=\linewidth]{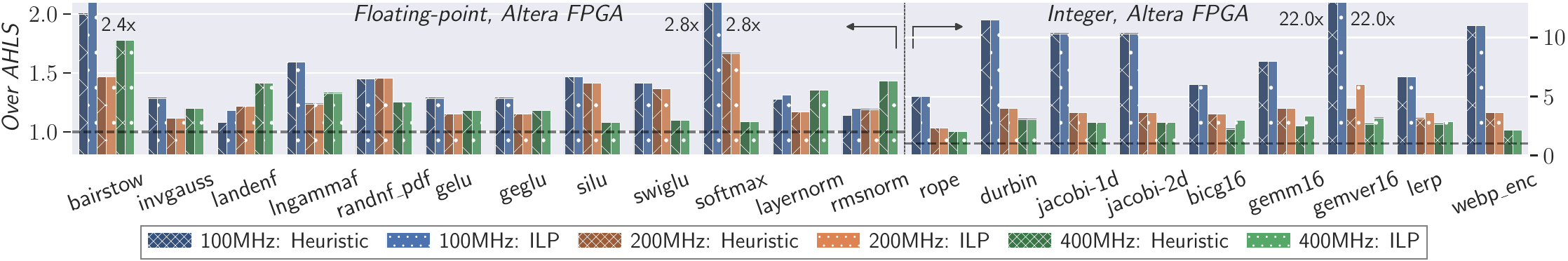}
  \caption{Speedup of \skyegg over Vitis HLS (on Xilinx FPGA, top) and Altera HLS (on Altera FPGA, bottom).}
  \label{fig:speedup}
\end{figure*}

\boldparagraph{Benchmarks.} Benchmarks we use are shown in \autoref{tab:eval-cases}. For integer, we use six scientific computing kernels from PolyBench~\cite{l-n_pouchet_polybenchc_2016} with three of them 16\x unrolled (labeled with postfix 16) to reflect common HLS optimization scenarios, and three signal processing kernels from Vitis Library~\cite{amd_inc_xilinxvitis_libraries_2025}. For float computation, we use five mathematics kernels from liquid-dsp~\cite{gaeddert_joseph_d_and_others_liquid-dsp_2025} and seven neural network kernels from ggml~\cite{georgi_gerganov_ggml_2025}. We additionally test \skyegg on synthetic benchmarks with 100-600 operations in both data types.

\boldparagraph{Mapping choices.}
Our evaluation employs mapping libraries covering both integer and floating-point operations. Integer arithmetic mappings include: LUT-based operations specified through RTL with synthesis directive (\texttt{use\_dsp} and \texttt{multstyle}) to prevent DSP inference, DSP48E2 or tennm\_mac slices instantiated as primitives, and vendor's divider IP cores. Floating-point operations utilize vendor's floating-point IP cores supporting various mathematical functions with configurable pipeline stages and resource usage. These libraries are aligned with vendor HLS tools' possible mapping sources.

\boldparagraph{Methodology.} We use HECTOR~\cite{xu_hector_2022} as the frontend and perform HLS for each basic block. We then run FPGA synthesis and fitting using vendor's toolchain and collect latency from HLS scheduling reports as well as timing slack and resource utilization from fitting reports. Overall latency is computed as the sum of all basic block latencies. We report speedup as $(\text{L}_{\text{Vendor}}+1)/(\text{L}_{\skyegg}+1)$ and compare timing closure and resource usage.

\subsection{Comparison with Commercial HLS Toolchains}
\label{sec:evaluation-base}

\autoref{fig:speedup} presents the performance comparison of \skyegg with Vitis HLS (top) and Altera HLS (bottom), using the exact ILP solver and the heuristic solver. Across all benchmarks and frequency targets, \skyegg consistently outperforms traditional sequential optimization strategies, achieving an average speedup of 3.12\x over Vitis HLS and 3.18\x over Altera HLS. Notably, the heuristic solver produced comparable results with exact ILP solving across, achieving the average speedup of 3.08\x and 3.09\x.

\subsubsection{Performance on Xilinx FPGA}
For floating-point cases, \skyegg achieves speedups ranging from 2.78\x to 5.22\x, with an average of 3.64\x.
\skyegg delivers an average speedup of 3.61\x at 100MHz and still achieves 2.42\x at the challenging 400MHz target. Because 400MHz approaches the maximum frequency of DSP48E2 and limits operation chaining, speedup at this target primarily comes from better mapping choices. We observed several mappings with different latencies but the same achievable frequency. In \texttt{rmsnorm}, square-root mappings from 10--13 cycles all reach 431MHz. \skyegg automatically selects the minimum 10-cycle variant, while Vitis HLS defaults to a conservative 28 cycles, reducing design latency from 73 to 25 cycles. We also see that more DSP usage does not necessarily improve performance. In \texttt{lngammaf}, logarithm mappings with \textit{"fulldsp"} in IP settings achieve 541MHz at 17 cycles, while \textit{"meddsp"} reaches the same frequency with 10 cycles using fewer DSP blocks. Traditional HLS tools with coarse models cannot handle these configuration decisions correctly, while \skyegg leverages profiled timing properties to select optimal mappings.

For integer cases, \skyegg achieves an average speedup of 2.38\x (up to 4.33\x). Lower frequencies yield higher speedups as \skyegg aggressively chains operations within single cycles due to better mappings, while Vitis HLS conservatively assigns each operation to separate cycles regardless of frequency targets. At 400MHz, the performance gap narrows as configurations become constrained. Despite this limited optimization scope, \skyegg consistently outperforms Vitis HLS at 400MHz.
The 1.0\x speedups in unrolled cases result from both tools selecting fully combinational implementations.

\begin{figure}[t]
\centering
\includegraphics[width=0.95\linewidth]{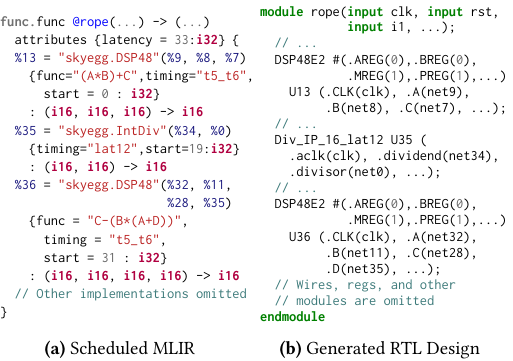}
\caption{Solution produced for the RoPE kernel.}
\label{fig:production}
\end{figure}

We use the \texttt{RoPE} kernel to illustrate a solution generated by \skyegg. \autoref{fig:production}(a) shows a representative subexpression and its selected mappings, including two DSP48E2 slices with configuration annotated. \autoref{fig:production}(b) shows the generated synthesizable Verilog. It directly instantiates configured DSP48E2 primitives and divider IP cores and inserts the required cut registers. Accompanying Tcl scripts configure the IP cores. The resulting design is verified to operate over 112MHz, satisfying the 100MHz design target.

\subsubsection{Performance on Altera FPGA}
\skyegg's exact ILP achieves an average speedup of 3.18\x over Altera HLS, with the heuristic solver achieving 3.09\x. For floating-point cases, speedups range from 1.08\x to 2.80\x, with an average of 1.37\x. Altera’s floating-point IPs are highly timing-conservative, resulting in substantial slack at low frequencies while offering no lower-latency options. Therefore, performance improvements primarily rely on algebraic transformation and operator chaining. For example, in \texttt{bairstow}, by rewriting \texttt{c-a*b} as \texttt{(-a)*b+c}, this whole expression can be fit into one FMA unit, and reduce the kernel latency to 6 cycles from 14.
\skyegg also mixes hard (DSP-based) and soft (LUT-based) implementations by scheduling position, while Altera HLS always prefers the hard DSP variant if possible, consuming more DSPs than \skyegg.
For integer cases, \skyegg achieves an average speedup of 5.49\x, up to 22.0\x. Speedups again decrease with frequency, from 10.09\x at 100MHz to 2.68\x at 400MHz.
In \texttt{bicg16}, Altera HLS serializes the reduction as independent 2-cycle multipliers followed by 1-cycle adders, while \skyegg correctly performs expression balancing and packs paired products into \textit{sumof2} mode DSPs (sum of two multipliers), cutting the kernel latency from 5 to 0.

\begin{table*}[t]
\centering
\small
\setlength{\tabcolsep}{4pt}
\setlength{\aboverulesep}{0pt}
\setlength{\belowrulesep}{0pt}
\setlength{\abovetopsep}{0pt}
\setlength{\belowbottomsep}{0pt}
\renewcommand{\arraystretch}{1.0}
\caption{Resource utilization and timing closure by vendor and frequency target. Resource usage is relative to the vendor's HLS tool in geometric-mean. T$_{\mathrm{S}}$ and T$_{\mathrm{V}}$ indicate timing met for \skyegg and the vendor's HLS tool, respectively.}
\label{tab:resource-vendor}
\begin{tabular}{c|c|ccccccccc|ccccccccc}
\toprule
\multirow{3}{*}{Vendor} & \multirow{3}{*}{Freq} & \multicolumn{9}{c}{Integer} & \multicolumn{9}{c}{Float} \\
\cmidrule(lr){3-11}\cmidrule(lr){12-20}
 & & \multicolumn{4}{c}{Heuristic} & \multicolumn{4}{c}{ILP} & \multirow{2}{*}{T$_{\mathrm{V}}$} & \multicolumn{4}{c}{Heuristic} & \multicolumn{4}{c}{ILP} & \multirow{2}{*}{T$_{\mathrm{V}}$} \\
\cmidrule(lr){3-6}\cmidrule(lr){7-10}\cmidrule(lr){12-15}\cmidrule(lr){16-19}
 & & FF & LUT & DSP & T$_{\mathrm{S}}$ & FF & LUT & DSP & T$_{\mathrm{S}}$ & & FF & LUT & DSP & T$_{\mathrm{S}}$ & FF & LUT & DSP & T$_{\mathrm{S}}$ & \\ \midrule
\multirow{3}{*}{Xilinx} & 100MHz & \textcolor{darkspringgreen}{0.64} & \textcolor{darkspringgreen}{0.71} & \textcolor{brickred}{1.15} & \dcheck & \textcolor{darkspringgreen}{0.72} & \textcolor{brickred}{1.12} & 1.06 & \dcheck & \dcheck & \textcolor{brickred}{1.53} & \textcolor{brickred}{1.70} & \textcolor{darkspringgreen}{0.27} & \dcheck & \textcolor{brickred}{1.34} & \textcolor{brickred}{1.70} & \textcolor{darkspringgreen}{0.32} & \dcheck & \dcheck \\
 & 200MHz & \textcolor{darkspringgreen}{0.64} & \textcolor{darkspringgreen}{0.67} & \textcolor{brickred}{1.15} & \dcheck & \textcolor{darkspringgreen}{0.58} & 1.09 & \textcolor{darkspringgreen}{0.81} & \dcheck & \dcheck & 0.91 & \textcolor{brickred}{1.55} & \textcolor{darkspringgreen}{0.45} & \dcheck & \textcolor{darkspringgreen}{0.83} & \textcolor{brickred}{1.53} & \textcolor{darkspringgreen}{0.47} & \dcheck & \dcheck \\
 & 400MHz & \textcolor{darkspringgreen}{0.78} & \textcolor{darkspringgreen}{0.86} & \textcolor{brickred}{1.15} & \dcheck & \textcolor{darkspringgreen}{0.78} & \textcolor{darkspringgreen}{0.86} & \textcolor{brickred}{1.15} & \dcheck & \dcheck & \textcolor{darkspringgreen}{0.89} & \textcolor{brickred}{1.25} & \textcolor{darkspringgreen}{0.56} & \dcheck & \textcolor{darkspringgreen}{0.88} & \textcolor{brickred}{1.26} & \textcolor{darkspringgreen}{0.56} & \dcheck & \dhalfcheck\ \textcolor{brickred}{(10/12)} \\
\midrule
\multirow{3}{*}{Altera} & 100MHz & \textcolor{darkspringgreen}{0.13} & \textcolor{darkspringgreen}{0.61} & \textcolor{darkspringgreen}{0.77} & \dcheck & \textcolor{darkspringgreen}{0.13} & \textcolor{darkspringgreen}{0.84} & \textcolor{darkspringgreen}{0.37} & \dcheck & \dcheck & 1.05 & \textcolor{brickred}{1.89} & \textcolor{darkspringgreen}{0.74} & \dcheck & 1.03 & \textcolor{brickred}{1.77} & \textcolor{darkspringgreen}{0.75} & \dcheck & \dcheck \\
 & 200MHz & \textcolor{darkspringgreen}{0.30} & \textcolor{darkspringgreen}{0.70} & \textcolor{darkspringgreen}{0.56} & \dcheck & \textcolor{darkspringgreen}{0.34} & 0.95 & \textcolor{darkspringgreen}{0.25} & \dcheck & \dcheck & \textcolor{darkspringgreen}{0.85} & 0.99 & \textcolor{darkspringgreen}{0.89} & \dcheck & 0.92 & \textcolor{brickred}{1.12} & \textcolor{darkspringgreen}{0.85} & \dcheck & \dcheck \\
 & 400MHz & \textcolor{darkspringgreen}{0.44} & \textcolor{darkspringgreen}{0.73} & \textcolor{darkspringgreen}{0.55} & \dcheck & \textcolor{darkspringgreen}{0.44} & \textcolor{darkspringgreen}{0.80} & \textcolor{darkspringgreen}{0.45} & \dcheck & \dhalfcheck\ \textcolor{brickred}{(1/9)} & 0.91 & 0.92 & 0.99 & \dcheck & 0.92 & 0.94 & 0.99 & \dcheck & \dcheck \\
\bottomrule
\end{tabular}
\end{table*}

\subsubsection{Resource and Timing Analysis.} As \autoref{tab:resource-vendor} shows, both \skyegg's solvers meet the timing in all designs, whereas both vendor tools have some failure cases at 400MHz. \skyegg's FF usage averages 0.5\x and 1.0\x of the vendor tools for integer and floating-point kernels, respectively. Floating-point FF usage is higher at low frequencies (1.5\x on Xilinx and 1.1\x on Altera) as \skyegg uses more soft mappings. LUT and DSP usage exhibit a frequency-dependent tradeoff. DSP slices are important but scarce resources in FPGAs, with LUTs outnumbering DSP slices by roughly 120 to 1 in the Xilinx XCKU3P FPGA~\cite{amd_inc_kintex_2025}. They are therefore most valuable when assigned to critical-path mappings. At low target frequencies, extra timing headroom allows \skyegg to map floating-point operations that gain no latency from DSP implementations to LUTs. At 100MHz, this results in 1.7\x LUT and 0.3\x DSP usage on Xilinx, and up to 1.9\x LUT and 0.7\x DSP usage on Altera, relative to the vendor tools. For integer kernels, \skyegg packs surrounding logic into selected DSP slices through fused modes, with LUT and DSP usage averaging between 0.7\x and 0.9\x of the vendor tools across both solvers. At high target frequencies, reduced timing headroom rules out more soft implementations and requires hard DSP mappings, narrowing the LUT gap.

In our evaluation, we assume that on-chip resources are sufficiently available and improve performance at a moderate cost. For severely resource-constrained scenarios, however, \skyegg's formulation can be extended with resource modeling to balance performance and resource usage as follows:
\begin{align}
  \forall k, \sum_{i} r_{m_i,k} \cdot b_{m_i} \leq R_k \label{eq:resource-constraint}
\end{align}
where $r_{m_i,k}$ denotes the amount of resource type $k$ consumed by the mapping $m_i$ and $R_k$ the available on-chip capacity. The heuristic solver could draw on established list-scheduling techniques~\cite{graham_bounds_1966} to solve such problems efficiently.

\subsection{Comparison with MapBuf}

\begin{figure}[t]
  \centering
  \includegraphics[width=\linewidth]{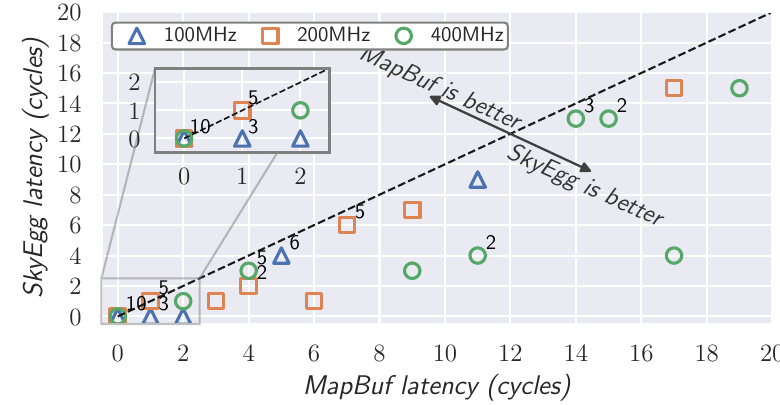}
  \caption{Latency comparison between MapBuf and \skyegg per benchmark component, colored by frequency target. Numbers annotate coincident component counts.}
  \label{fig:mapbuf}
\end{figure}

MapBuf\cite{wang_mapbuf_2023} jointly optimizes LUT mapping and buffer insertion in HLS-produced dataflow circuits over gate-level netlists. We compare \skyegg with MapBuf on DSP mapping tasks targeting Xilinx FPGA. We extend MapBuf by lifting mapping to operator-level DFGs, generalizing \textit{cuts} to support operator fusion and multicycle mappings, and replacing its depth-based delay estimation with \skyegg's timing model.

\autoref{fig:mapbuf} shows the result. \skyegg consistently outperforms MapBuf, achieving a 1.37\x speedup on average. The speedup primarily comes from equality saturation, which exposes fused-hardware mappings beyond structures already present in MapBuf's input DFG. Moreover, \skyegg separates algebraic transformations from hardware mappings, eliminating dedicated generators for individual fusion patterns and allowing easy extension without changing the optimization formulation. For resource usage, \skyegg's designs on average use 0.87\x of MapBuf's LUTs, 0.85\x of its FFs, and 1.06\x of its DSPs. Overall, by better exploiting the computational capabilities of DSP blocks through their pre-adders and post-multiply ALUs, \skyegg achieves higher performance and lower LUT usage with comparable DSP usage.

\subsection{Scalability Analysis}
\label{sec:scalability}

\begin{figure}[t]
  \centering
  \includegraphics[width=\linewidth]{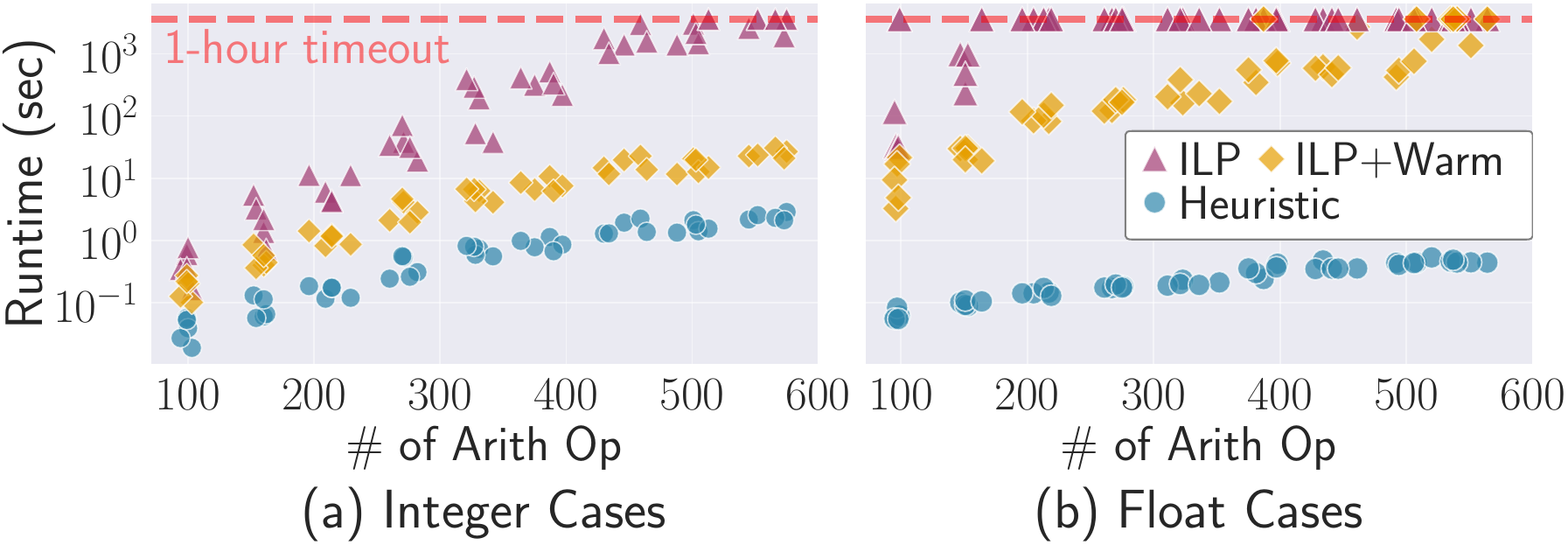}
  \caption{Scheduler runtime on synthesized cases. ILP+Warm uses the heuristic schedule as a warm start.}
  \label{fig:scalability}
\end{figure}

We evaluate the runtime of the exact ILP solver and the scalable heuristic solver on synthetic integer and floating-point benchmarks containing 100 to 600 arithmetic operations, as shown in \autoref{fig:scalability}. For integer cases, the heuristic solver completes within three seconds, whereas the standalone ILP solver times out in 4 of 9 cases exceeding 500 operations. Floating-point operators offer richer configuration choices, further expanding the design space. The standalone ILP solver completes only 8 of 45 cases within the one-hour limit, whereas the heuristic solver completes all 45 cases in under a second. These heuristic schedules also serve as warm starts, enabling ILP to solve additional large cases, including cases beyond 400 floating-point operations. Regardless, the heuristic incurs less than 10\% latency overhead relative to ILP across all cases completed by both solvers. Together, these results demonstrate the practicality of our method.

\subsection{Ablation Study}
\label{sec:ablation}

\begin{figure}[t]
  \centering
  \includegraphics[width=\linewidth]{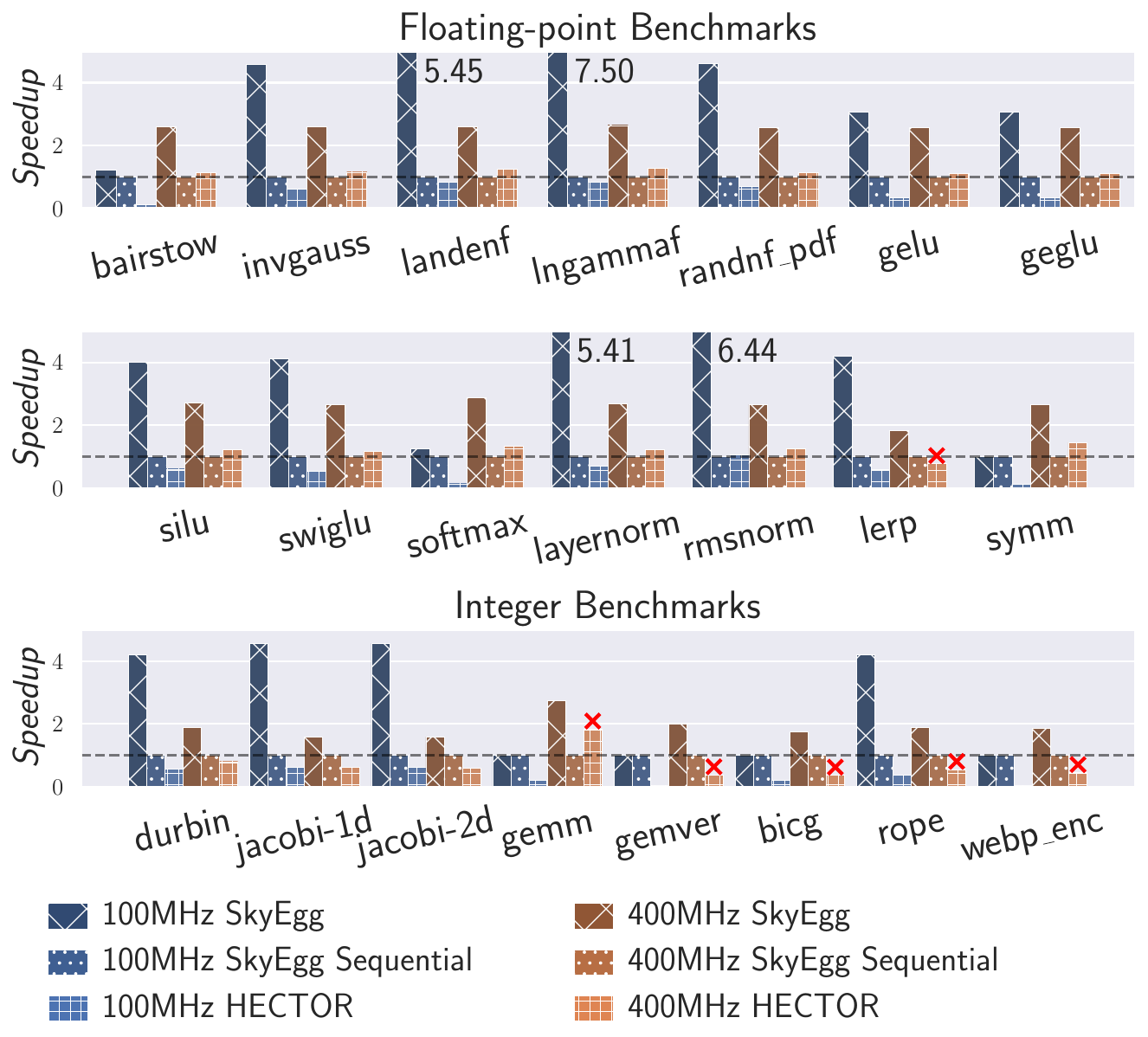}
  \caption{Comparison of latency across different ablation settings. Speedup is computed relative to \texttt{SkyEgg Sequential}. Red cross markers (\textcolor{red}{$\times$}) denote timing unmet.}
  \label{fig:ablation}
\end{figure}

We further perform an ablation study to isolate the impact of transformation- and mapping-aware scheduling.
We compare \skyegg against two baselines: \texttt{SkyEgg Sequential}, which disables algebraic transformations and uses a sequential flow that selects minimum-latency mappings satisfying timing constraints before scheduling, and \texttt{HECTOR}, which uses fixed mapping choices regardless of target frequency.

\autoref{fig:ablation} shows the results on the same benchmarks from \autoref{sec:evaluation-base} at 100MHz and 400MHz. Compared to the sequential approach, \skyegg achieves an average speedup of 3.52\x and 2.34\x, respectively. This improvement demonstrates the benefit of preserving algebraic alternatives while selecting hardware mappings and schedules together. These coordinated decisions enable better mappings and operation chaining than the sequential baseline. Consistent with previous comparisons, \skyegg achieves larger gains at lower frequencies. At higher frequencies, increasing DSP usage makes suboptimal mapping assumptions more costly. Compared to the fixed mapping approach, \skyegg achieves up to 20.0\x speedup at 100MHz. At 400MHz, half of the benchmarks under the fixed-mapping approach fail to meet timing constraints, while \skyegg generates timing-feasible designs by leveraging accurate mapping latency. Fixed mapping is also common in open-source HLS tools such as Dynamatic~\cite{josipovic_dynamically_2018}. \skyegg can complement these tools by adapting mapping and scheduling decisions to target timing requirements.

\section{Related Work}
\label{sec:related-work}

\boldparagraph{E-graph Techniques for Hardware Synthesis.}
E-graphs and equality saturation have been applied at several levels of hardware synthesis.
For source and datapath optimization, SEER~\cite{cheng_seer_2024} performs loop and datapath super-optimization at the program level, while \cite{coward_automatic_2022} and Chassis~\cite{saiki_target-aware_2025} optimize datapaths through rewrite rules. IMpress~\cite{ustun_impress_2022} explores resource-balanced decompositions of large integer multiplications before HLS.
Most of these approaches do not incorporate target-specific hardware models and extract one best expression before scheduling, discarding alternatives that may yield better performance in subsequent stages.
E-Syn~\cite{chen_e-syn_2024} and E-morphic~\cite{chen_e-morphic_2025} apply equality saturation to logic-level technology mapping for area and delay optimization, while EqMap~\cite{hofmann_eqmap_2025} applies it to FPGA LUT remapping.
Churchroad~\cite{smith_scaling_2024} combines equality saturation with program synthesis to discover efficient mappings of RTL expressions onto DSP blocks.
\skyegg instead represents algebraic transformations and hardware mappings as rewrite-rules within a unified e-graph, allowing each space to be extended independently.

\boldparagraph{Hardware-Aware HLS Scheduling.}
Prior work couples HLS optimization with hardware concerns at different levels of abstraction.
AutoBridge~\cite{guo_autobridge_2021}, TAPA~\cite{guo_tapa_2023}, and RapidStream~\cite{guo_rapidstream_2022, lau_rapidstream_2024} co-optimize HLS with physical design through floorplanning and pipelining, primarily addressing inter-module interconnect and multi-die partitioning.
\cite{cabodi_speeding-up_2010, ustun_accurate_2020} incorporate resource and operation-delay estimates, respectively, while ~\cite{ye_subgraph_2024, zheng_fast_2014, rizzi_iterative_2023} leverage logic-synthesis, placement-and-routing, or mapping feedback to refine HLS schedules.
\cite{tan_mapping-aware_2015, wang_mapbuf_2023} more directly couple scheduling with logic mapping or buffer insertion, but remain limited to LUT-level logic.
Despite these different forms of hardware awareness, all operate on a fixed algebraic structure and cannot evaluate equivalent forms together with their mappings and schedules.
\skyegg addresses this gap through heterogeneity-aware synthesis, preserving algebraic transformations and configurable mappings through scheduling to coordinate all three decisions.

\section{Conclusion}

This work presents \skyegg, a framework for heterogeneity-aware hardware synthesis via equality saturation. \skyegg encodes algebraic transformations, hardware mappings, and their configurations as rewrite rules and formulates a joint design space which optimizes transformation, mapping, and scheduling together. It also provides a scalable heuristic solver.
\skyegg achieves average speedups of 3.12\x over Vitis HLS and 3.18\x over Altera HLS across target frequencies, showing the end-to-end benefits of its unified optimization.

\bibliographystyle{ACM-Reference-Format}
\bibliography{skyegg}

\end{document}